\documentclass[twocolumn,usenatbib]{mnras2}
\pdfoutput=1 
\usepackage{amsmath,amstext}
\usepackage[T1]{fontenc}
\usepackage{amssymb}
\input{epsf}
\usepackage{graphicx}
\usepackage{ae,aecompl}

\usepackage{hyperref}
\usepackage{amsmath}
\usepackage{amssymb}
\usepackage{mathtools}
\usepackage{bm}
\usepackage{cleveref}
\usepackage{tensor}
\usepackage{braket}
\usepackage{mhchem}

\usepackage{color}

\newcommand{\spc}{\quad \quad \quad}

\def\be{\begin{equation}}
\def\ee{\end{equation}}
\def\beq{\begin{eqnarray}}
\def\eeq{\end{eqnarray}}

\title[Invariance of the entropy]{Proving the Lorentz invariance of the entropy and the covariance of thermodynamics}
\author[L.~Gavassino]{L.~Gavassino\\
Nicolaus Copernicus Astronomical Center, Polish Academy of Sciences, ul. Bartycka 18, 00-716 Warsaw, Poland}
\begin{document}
\maketitle

\begin{abstract}
The  standard  argument  for  the  Lorentz  invariance  of  the  thermodynamic  entropy in equilibrium is based on the assumption that it is possible to perform an adiabatic transformation whose only outcome is to accelerate a macroscopic body, keeping its rest mass unchanged. The validity of this assumption constitutes the very foundation of relativistic thermodynamics and needs to be tested in greater detail. We show that, indeed, such a transformation is always possible, at least in principle. The only two assumptions invoked in the proof are that there is at least one inertial reference  frame  in  which  the  second  law of thermodynamics is  valid  and  that  the  microscopic theory describing the internal dynamics of the body is a field theory, with Lorentz invariant Lagrangian density. The proof makes no  reference  to  the  connection  between  entropy  and  probabilities  and  is  valid  both within classical and quantum physics. To avoid any risk of circular reasoning, we do not postulate that the laws of thermodynamics are the same in every reference frame, but we obtain this fact as a direct consequence of the Lorentz invariance of the entropy.
\end{abstract}


\section{Introduction} 

The total thermodynamic entropy $S$, in equilibrium, must be Lorentz invariant. Every statistical mechanical view on thermodynamics agrees on this point. Whether we identify $S$ with the Boltzmann entropy \citep{degroot_book,cercignani_book}, or with the Gibbs/Shannon entropy \citep{Jaynes1965,Adami2011}, or with the von Neumann entropy \citep{Israel_1981_review,hakim_book}, its Lorentz invariance seems inescapable. This fact is also a foundational feature of relativistic fluid dynamics \citep{Israel_Stewart_1979,rezzolla_book} and of thermal quantum field theory \citep{Becattini2016}. 

Intuitively, the invariance of the entropy with respect to Lorentz transformations is usually justified by invoking its statistical connection with microscopic probabilities (or numbers of quantum states), which are supposed to have an invariant nature \citep{Nakamura2012,PARVAN2019}. However, when it comes to proving rigorously, from first principles, that the \textit{thermodynamic} entropy (namely, the macroscopic state function which is subject to the second law) must necessarily be a scalar, some conceptual problems arise and it is easy to fall into circular reasoning.

The thermodynamic argument for the Lorentz invariance of the entropy that is often repeated in the literature \citep{Farias2017,Mares2017} is an oversimplified version of an argument originally proposed by \cite{Planck_1908}. Consider the following thought experiment: a body $X$ is accelerated from being at rest with respect an observer $A$ to being at rest with respect to an observer $B$ (in motion with respect to $A$). If the process is adiabatic, it is reversible, hence the entropy of $X$ measured by $A$ is the same before and after the acceleration: $S_{i}(A)=S_f(A)$. Now let's assume that during this process the rest-frame properties of the body do not change (hence we may call this process a \textit{pure acceleration}). It follows that the initial state, as seen by $A$, is identical to the final state, as seen by $B$, which implies $S_i(A)=S_f(B)$ (recall that the entropy is a state function). Thus, $A$ and $B$ agree on the value of the entropy at the end of the process, $S_f(A)=S_f(B)$, proving the Lorentz invariance of the entropy. 

The problem with this argument is that what determines whether a process is reversible or not is the difference in entropy between the initial and the final state (if $\delta S=0$, the process is reversible). Hence, assuming that pure accelerations are reversible is equivalent to assuming that the entropy does not depend on the velocity of the body, which is exactly what we are trying to prove. To the best of our knowledge, the first author who noted this circularity problem was \cite{vanKampen1968}, who elevated the existence of reversible pure accelerations to the rank of fundamental postulate of relativistic thermodynamics. He showed that no entirely thermodynamic argument can be used to prove ab initio the Lorentz invariance of the entropy, but, to set the foundations of covariant thermodynamics rigorously (and to avoid any circularity issue), one only needs to postulate that pure accelerations are reversible.

The goal of the present paper is to explore the validity of van Kampen's postulate in greater detail. In fact, from an operational point of view, the postulate can be rephrased as follows: \textit{adiabatic accelerations (i.e. slow variations of velocity generated by weak mechanical forces) do not alter the rest-frame properties of a body; in particular, they do not affect its rest mass}.  Given that this is a simple statement about the behaviour of many-particle systems subject to external forces, it should be possible to test it using relativistic dynamics and quantum field theory.

We remark that the purpose of this paper is not to convince the reader that the entropy is Lorentz invariant; this is already a well established fact \citep{Israel2009_book}. Instead, the aim is to explain why this is the \textit{only} possibility and to prove that any alternative construction of relativistic thermodynamics would lead to serious inconsistencies.

Throughout the paper we adopt the signature $(-,+,+,+)$ and work with natural units $c=k_B=\hbar=1$. The study is performed within special relativity, hence the metric is assumed flat. Greek space-time indices $\mu,\nu,\rho$ run from 0 to 3, while latin space indices $j,k$ run from 1 to 3.

\section{The rationale of the argument}

If we want to make our argument solid and unquestionable, we need, first, to understand which assumptions about relativistic thermodynamics we are reasonably allowed to uphold, and which might lead us to circular reasoning.

\subsection{Must the laws of thermodynamics be the same in every reference frame?}

It is possible to formulate many arguments for the Lorentz invariance of the entropy, based on the assumption that the laws of thermodynamics should be the same in every reference frame. A well-known example is Planck's original argument (which is more refined than the version reported in the introduction), of which we present a slightly more formal version in appendix \ref{AAA}. The rationale of Planck's argument \citep{Planck_1908}, and of most of the other thermodynamic arguments present in the literature, is that the entropy is ultimately a rule, which dictates which processes are possible (for thermally isolated systems) and which are not. For example, if a macroscopic state $\psi$ has a lower entropy than a macroscopic state $\psi'$, this means that, if we keep the system thermally isolated, the process $\psi \rightarrow \psi'$ is possible, while the inverse process is not. Clearly, statements about the possibility for a process to occur cannot depend on the reference frame, hence the entropy must be Lorentz invariant.

The problem with these arguments is that they all treat thermodynamics as a fundamental theory, which should be subject to the principle of relativity in the same way as dynamics is, and whose laws should, therefore, be equally valid in every reference frame. In other words, it is assumed in these arguments that thermodynamics should share the same symmetries of dynamics. However, we already know that there is at least one symmetry for which this is not true: CPT. While CPT is a fundamental symmetry in quantum field theory \citep{Weinberg_QFT_1995}, it is manifestly violated by the second law of thermodynamics. This shows us that we are in general not allowed to treat thermodynamics on the same footing as dynamics. 

The fundamental distinction between dynamics and thermodynamics is that dynamics studies the evolution of systems with arbitrary initial conditions, which implies that the solutions of the equations which govern dynamics form a set $\Lambda$ that is necessarily invariant under the action of the symmetry group $\mathcal{G}$ of the spacetime ($\mathcal{G}\Lambda = \Lambda$).
On the other hand, thermodynamics deals only with a subset $\lambda \subset \Lambda$ of solutions, whose initial conditions have precisely those statistical properties (e.g. molecular chaos, see \citealt{huang_book}) which give rise to the second law as an emergent quality. It might be the case (and for CPT it is the case!) that these constraints on the initial conditions lead to a symmetry breaking, namely to a situation in which $\mathcal{G} \lambda \neq \lambda$. Considering that specifying the laws of thermodynamics is essentially equivalent to specifying $\lambda$, it follows that thermodynamics might in turn not be symmetric under $\mathcal{G}$. 

Let us remark that we are not claiming that the laws of thermodynamics are not Lorentz covariant. They are. But (as we will show in subsection \ref{vanKampennPost}) their covariance follows from the invariance of the entropy, and not vice-versa. Thus, in a paper whose goal is to prove the invariance of the entropy, we are not allowed to include the assumption that thermodynamics is the same in every reference frame among the hypotheses. 

As a last comment on this issue, we point out that, if one adopts Jaynes' statistical justification for the second law \citep{Jaynes1965}, then the initial conditions that give rise to $\lambda$ are actually the overwhelming majority of initial conditions which are compatible with the initial macroscopic data (in the thermodynamic limit). Hence, it is to be expected that, if the group $\mathcal{G}$ conserves the \textit{causal} ordering of the events (namely if it does not convert initial states into final states), then $\lambda$ should be approximately invariant under $\mathcal{G}$. This would explain why thermodynamics is not invariant under CPT (namely $\text{CPT}  \lambda \neq \lambda$) while it is expected to be invariant under the proper orthochrounus Lorentz group ($SO^+(3,1) \lambda = \lambda$). In fact, CPT converts initial data into final data, whereas $SO^+(3,1)$ conserves the causal structure of the field equations by construction \citep{Peskin_book}. This is the actual statistical justification for the covariance of thermodynamics, because it is not grounded on the interpretation that one chooses to give to the entropy, but on the statistical origin of irreversibility, which constitutes the very foundation of thermodynamics. However, as this argument is qualitative, and thermodynamics does not entirely reduce to Jaynes' view \citep{Rigol2008,Gogoglin2016}, it is important to have also a more formal proof, which is the purpose of the present paper.

\subsection{The assumptions of the argument}\label{assumptionsSSSSS}

Motivated by the complication outlined in the previous subsection, we need to make an argument for the Lorentz invariance of the entropy which does not build on the assumption that the second law of thermodynamics is valid for every observer. Instead, we will base our argument only two uncontroversial assumptions, namely
\begin{enumerate}
\item[(i) -] There is a global inertial reference frame $A$ in which it is possible to unambiguously define a notion of entropy $S$ that obeys the second law: $\dot{S} \geq 0$. In this reference frame, bodies may interact with each other, accelerate, decelerate and be destroyed, but the total entropy of isolated systems can never decrease.
\item[(ii) -] The microscopic dynamics can be modelled using a field theory, governed by a Lorentz invariant Lagrangian density.
\end{enumerate}
Assumption (i) is simply the requirement that there is \textit{at least} one observer for which the laws of thermodynamics, in their standard  ``textbook'' formulation, are valid. Assumption (ii) is the statement that, although thermodynamics might in principle not admit a covariant formulation, dynamics does. We are enforcing the principle of relativity on the underlying microscopic theory, rather than imposing it directly on thermodynamics.

Throughout the rest of this paper, we will always work in the reference frame $A$ introduced in assumption (i), so that thermodynamics works as usual. In this way we will avoid any possible source of confusion.

\subsection{van Kampen's argument}\label{vanKampennPost}

Let us now briefly revisit van Kampen's argument for the Lorentz invariance of the entropy \citep{vanKampen1968}.

We consider an isolated (freely moving) body in thermodynamic equilibrium with total four-momentum $p^\nu$ and rest mass $M=\sqrt{-p^\nu p_\nu}$. The entropy in equilibrium must be a function of the constants of motion of the body. To capture the essence of the problem, we assume for simplicity that the only relevant constants of motion are the components of the four-momentum\footnote{The only relevant constants of motion of an ergodic body are four-momentum, angular momentum (tensor) and conserved charges (like the baryon number). If we work at fixed conserved particle numbers, and assume that the body is non-rotating, equation \eqref{laUno} follows. The volume cannot be treated as an independent variable in relativistic thermodynamics. In fact, if a given volume is imposed through external walls, the body is not isolated. Finite isolated bodies are self-bounded, hence their volume is an equilibrium property (like the volume of stars and nuclei) and not a free parameter \citep{GavassinoTermometri}.}, so that
\begin{equation}\label{laUno}
S=S(p^\nu).
\end{equation}
At this stage, the function $S(p^\nu)$ may be completely arbitrary, because (as we anticipated) we are not excluding a priori the possibility that thermodynamics may break Lorentz covariance. Similarly to what we did in the introduction, let us postulate that it is possible to make infinitesimal \textit{reversible pure accelerations}, namely transformations $\delta p^\nu$ such that
\begin{equation}\label{2}
\delta S = \dfrac{\partial S}{\partial p^\nu} \delta p^\nu =0 \spc (\text{reversible})
\end{equation}
and
\begin{equation}\label{3}
\delta M = -\dfrac{p_\nu}{M} \delta p^\nu =0 \spc (\text{pure acceleration}).
\end{equation}
If these accelerations can have arbitrary direction (i.e. if those $\delta p^\nu$ that satisfy \eqref{2} and \eqref{3} form a 3D plane), then it follows that there is a function $T$ such that
\begin{equation}
dS = \dfrac{dM}{T},
\end{equation}
which in turn implies
\begin{equation}
S=S(M).
\end{equation}
The fact that the entropy can be written as a function of a Lorentz scalar implies that, when we perturb the system, the second law of thermodynamics ($\dot{S} \geq 0$) takes the form of a Lorentz-invariant statement:
\begin{equation}
\dfrac{\dot{M}}{T(M)} \geq 0.
\end{equation}
But this implies that the set $\lambda$ of all the initial conditions which realise the second law is invariant under the action of the proper orthochronous Lorentz group (formally, $SO^+(3,1) \, \lambda = \lambda$), proving that thermodynamics admits a covariant formulation, in which $S$ is a Lorentz scalar. This sets solid foundations for relativistic thermodynamics.

Our goal, now, is to prove that, if assumptions (i) and (ii), as stated in the previous subsection, are valid, a set of infinitesimal transformations that satisfy both \eqref{2} and \eqref{3} always exist (at least in principle), converting van Kampen's postulate into a theorem.

\section{Reversible accelerations}

Our first task is to understand how we may induce an ideal reversible acceleration on a body. Following \cite{landau5}, the most perfect form of reversible process is an \textit{adiabatic} process, namely an infinitely slow transformation in which the system is kept thermally isolated. Such processes can be modelled, at the microscopic level, as transformations induced by a weak and slow time-dependence of the microscopic Hamiltonian. Our aim is to design an adiabatic transformation which can alter the state of motion of a relativistic body. 

\subsection{Small kicks}

Let $\varphi_i$ be the microscopic fields of the body and $\mathcal{L}_{\text{Body}}(\varphi_i,\partial_\mu \varphi_i)$ the Lagrangian density governing the microscopic dynamics. Assume that we are able to generate and control an external potential $\phi$ (a real scalar field, for simplicity), which interacts with the body through a small dimensionless coupling constant $\epsilon$, so that the action takes the simple form 
\begin{equation}\label{Action}
\mathcal{I}[\varphi_i] = \int \bigg[ \mathcal{L}_{\text{Body}}(\varphi_i,\partial_\mu \varphi_i) + \epsilon \, \phi \, G(\varphi_i) \bigg] \, d^4 x \, ,
\end{equation}
where $G(\varphi_i)$ is an observable. The potential $\phi$ is an \textit{assigned} real function of the coordinates $\phi(x^\nu)$. It is not a dynamical degree of freedom of the total system (``$\, \text{body}+\phi \,$''), but it plays the role of a source in the action $\mathcal{I}[\varphi_i]$, which breaks the Poincar\'e invariance of the theory. In a quantum description, the field $\phi$ plays the role of a classical source \citep{Peskin_book}; it is not a quantum field. We model $\phi$ in this way because we want to treat it as a purely mechanical and non-statistical entity (like any other source of thermodynamic work, see e.g. \citealt{GavassinoTermometri}), so its evolution must be completely known and cannot be affected by the statistical fluctuations of the dynamical fields $\varphi_i$. In this sense, the potential $\phi$ may be seen as an analogue of the perfectly reflecting walls of an adiabatic box: it carries no entropy. This implies that the body remains thermally isolated \citep{landau5} and the second law of thermodynamics holds for the entropy of the body alone \citep{Jaynes1965}, also during its interaction with $\phi$.

Assume that $\phi=0$ for $t\leq 0$ (recall that we always work, for clarity, in the reference frame $A$ in which we have a notion of entropy). The configuration of the system for $t\leq 0$ is the initial state of the body, which is assumed to be an equilibrium state, with four-momentum $p^\nu$. At $t= 0$ we switch on the external potential and we keep it active for a finite time $\tau$, namely
\begin{equation}
\phi \neq 0  \quad \text{ for} \quad  t \in (0,\tau).
\end{equation} 
No assumption about the duration $\tau$ of the process, nor about the exact space-time dependence of $\phi(x^\nu)$, is made. We only require that there is at least a small region of space-time (between the times $0$ and $\tau$) in which 
\begin{equation}
(\partial_1 \phi)^2 + (\partial_2 \phi)^2 + (\partial_3 \phi)^2 >0 \, ,
\end{equation}
so that we know that the action \eqref{Action} is not invariant under space translations, breaking the Noether conservation of linear momentum of the body. At the end of the process ($t=\tau$), the four-momentum of the body has changed of a finite amount $\delta p^\nu$. After some more time passes, the system can reach a new state of equilibrium, whose entropy is $S(p^\nu+ \delta p^\nu)$. The total variation of entropy experienced by the system during all this process (including the final relaxation to a new equilibrium) is the finite difference
\begin{equation}
\delta S = S(p^\nu +\delta p^\nu) - S(p^\nu).
\end{equation}

The aforementioned process may be interpreted as a small kick generated by an ideal mechanical device:
\begin{itemize}
\item For $t\leq 0$ the body is completely isolated and in thermodynamic equilibrium. It moves freely across space-time, with initial mass $M=\sqrt{-p^\nu p_\nu}$ and center-of-mass four-velocity $u^\nu=p^\nu/M$. It is in the maximum entropy state possible (as measured in the frame $A$) compatible with this value of four-momentum.
\item For $0< t < \tau$ the body interacts with a mechanical device with no microscopic degrees of freedom (zero entropy). The interaction is mediated by a potential $\phi$, which is generated solely by the device (and therefore carries no entropy). Through this interaction, the body feels a force, which impresses on it a small kick, changing its total four-momentum by an amount $\delta p^\nu$. This amount of energy and momentum is transferred through $\phi$ to the device, which is however not explicitly modelled here. 
\item For $t \geq \tau$ the body is again completely isolated and has time to dissipate all the fluctuations and vibrations induced by the kick, to reach a new equilibrium.
\end{itemize}

Comparing this description with subsection 6.2 of our previous paper \citep{GavassinoTermometri}, one can see that the variation of four-momentum $\delta p^\nu$ produced in a kick has the nature of pure work (using the terminology we introduced there: $\delta p^\nu = \delta \mathcal{W}^\nu$), because the external agent can be modelled as a purely mechanical entity. Hence, kicks are the simplest form of work-type energy-momentum transfers in relativistic thermodynamics.

\subsection{Infinite infinitesimal kicks}\label{infinite infitesimal}

The key insight which leads us to a notion of adiabatic acceleration is how the changes $\delta p^\nu$ and $\delta S$ scale with the strength of the coupling constant $\epsilon$, in the limit in which $\epsilon \rightarrow 0$. We take this limit at fixed initial state of the body (for $t \leq 0$) and keep the function $\phi(x^\nu)$ fixed.

Since $\epsilon$ quantifies how strongly the system reacts to the presence of the external potential $\phi$ ($\epsilon$ is analogous to the coupling constant $q$ in the electrostatic force $\textbf{F}=q\textbf{E}$), it is easy to see that, to the leading order in $\epsilon$, we have the scaling
\begin{equation}\label{dpe}
\delta p^\nu \sim \epsilon \, .
\end{equation}
However, the variation of the entropy scales differently. In fact, the second law implies $\delta S(\epsilon) \geq 0$ $\forall \, \epsilon$. On the other hand, $\epsilon$ may have arbitrary sign\footnote{Nothing forbids us to impose $\epsilon <0$ in the action \eqref{Action}. In fact, changing the sign of $\epsilon$ keeping $\phi$ fixed is equivalent to keeping $\epsilon$ fixed and changing the sign of $\phi$.}, which implies that if we assume $\delta S \sim \epsilon$ we get a contradiction with the second law. Thus, the leading order must be
\begin{equation}\label{dse}
\delta S \sim \epsilon^2 \, ,
\end{equation}
or higher (but even).

Now, consider a sequence of $N$ kicks ($N \rightarrow +\infty$) with a coupling constant $\epsilon = 1/N \rightarrow 0$. The total variation of the four-momentum (due to the whole sequence of kicks) is
\begin{equation}
(\delta p^\nu)_{\text{N kicks}} \sim N \times (\delta p^\nu)_{\text{1 kick}} \sim N \times \dfrac{1}{N} =1 \, ,
\end{equation} 
while the total variation of entropy is
\begin{equation}
(\delta S)_{\text{N kicks}} \sim N \times (\delta S)_{\text{1 kick}} \sim N \times \dfrac{1}{N^2} = \dfrac{1}{N} \, .
\end{equation}
This implies that, as the number of kicks goes to infinity and their intensity goes to zero, the resulting transformation is non-trivial ($\delta p^\nu$ is finite) and reversible ($\delta S = 0$). Hence, we have just built a microscopic model for a reversible acceleration. As expected, it is infinitely slow (duration $\geq \tau \times N \rightarrow +\infty$), so we have rediscovered the well-established fact that adiabatic transfers of energy-momentum (i.e. infinitely slow processes in which $\delta p^\nu = \delta \mathcal{W}^\nu$) are reversible (see \citealt{GavassinoIorda2021}, section 6, for another example). Note also that the reversibility of this transformation has been justified using only condition (i), namely the second law of thermodynamics; no other property of the entropy has been invoked. 

In order to show that this reversible process is a \textit{pure} acceleration, which would prove van Kampen's postulate, see equation \eqref{3}, we only need to show from microphysics that necessarily
\begin{equation}\label{iltaskuzzo}
\delta M \sim \epsilon^2 \, ,
\end{equation}
as this would immediately imply that $(\delta M)_{\text{N kicks}} \sim 1/N \rightarrow 0$. The next two sections of the paper contain two alternative proofs of \eqref{iltaskuzzo}.

\section{Variation of the mass induced by a kick: field theory approach}

We derive equation \eqref{iltaskuzzo} from a field theory point of view. 

\subsection{Classical case}\label{classonz}

Let us define the tensor field 
\begin{equation}
T\indices{^\mu _\nu} = \mathcal{L}_{\text{Body}} \, \delta\indices{^\mu _\nu} -\dfrac{\partial \mathcal{L}_{\text{Body}}}{\partial(\partial_\mu \varphi_i)} \, \partial_\nu \varphi_i  \, ,
\end{equation}
where we are applying Einstein's summation convention also to the label $i$. Given that the Euler-Lagrange equations, computed from the action \eqref{Action}, are
\begin{equation}
  \dfrac{\partial \mathcal{L}_{\text{Body}}}{\partial \varphi_i} - \partial_\mu \bigg( \dfrac{\partial \mathcal{L}_{\text{Body}}}{\partial(\partial_\mu \varphi_i)} \bigg) = -\epsilon \, \phi \, \dfrac{\partial G}{\partial \varphi_i} \, ,
\end{equation}
one can easily show that $T\indices{^\mu _\nu}$ obeys the equation
\begin{equation}\label{nonConse}
\partial_\mu T\indices{^\mu _\nu} = -\epsilon \, \phi \, \partial_\nu G.
\end{equation}
This implies that for $t \leq 0$ and $t \geq \tau$, i.e. in those space-time regions in which $\phi=0$, the tensor field $T\indices{^\mu _\nu}$ is conserved, namely $\partial_\mu T\indices{^\mu _\nu} =0$. Indeed, $T\indices{^\mu _\nu}$ is the Noether stress energy tensor associated with $\mathcal{L}_{\text{Body}}$ \citep{Weinberg_QFT_1995}, therefore it can be used to define the four-momentum of the body before and after the kick, by means of the formulas
\begin{equation}
\begin{split}
& p_\nu = \int T\indices{^0 _\nu} \, d^3 x  \spc \quad \, \, \, \,  \text{for }t \leq 0 \, \text{ (before the kick)} \\
& p_\nu + \delta p_\nu = \int T\indices{^0 _\nu} \, d^3 x  \quad \quad \text{for }t \geq \tau \, \text{ (after the kick)} \, . \\
\end{split}
\end{equation}
Recalling that the four-velocity of the center of mass is $u^\nu =p^\nu/M$ and applying Gauss' theorem to the spacetime region $\mathcal{R}=(0,\tau)\times \mathbb{R}^3$ (assuming that the body is finite, so that the fields are zero at infinity), one can use \eqref{nonConse} to prove that
\begin{equation}\label{hisguz}
    \delta M = -u^\nu \delta p_\nu = \epsilon \int_{\mathcal{R}}   \phi \, u^\nu \partial_\nu G \, d^4 x \, .
\end{equation}
The second equality in equation \eqref{hisguz} is exact, whereas the first is valid up to the first order in $\epsilon$. In the limit of small $\epsilon$, we may use linear response theory and model $G$ as the sum
\begin{equation}\label{bartycka}
    G=G_0 + \epsilon \, G_1 \, ,
\end{equation}
where $G_0$ is the value that the observable $G(\varphi_i)$ would have (on the spacetime point under consideration) if no kick were impressed on the body, while $\epsilon \, G_1$ describes the perturbation to $G$ due to the kick. Let us focus on the function $G_0(x^\nu)$. If no kick were impressed on the body, the body would remain in a state of thermodynamic equilibrium, and would be drifting rigidly with constant four-velocity $u^\nu$ without experiencing any macroscopic deformation, because it would keep the equilibrium shape. This implies that statistically (i.e. once we average over the microscopic fluctuations) we must have 
\begin{equation}\label{worldlinecons}
    u^\nu \partial_\nu G_0=0.
\end{equation} 
This formula can be justified with the qualitative argument above, but it can also be proved rigorously from condition (ii), see appendix \ref{Rigiduz}. If we plug \eqref{bartycka} into \eqref{hisguz}, we obtain
\begin{equation}
   \delta M = -u^\nu \delta p_\nu =  \epsilon^2 \int_{\mathcal{R}}   \phi \, u^\nu \partial_\nu G_1 \, d^4 x  \sim \epsilon^2  \, ,
\end{equation}
which is what we wanted to prove (see equation \eqref{iltaskuzzo} and recall that $\epsilon^2 = 1/N^2$). In conclusion, van Kampen's postulate is valid, the entropy is Lorentz invariant and thermodynamics admits a covariant generalization.

\subsection{Quantum case}

The above calculations are essentially the same if we move to a quantum context. Equation \eqref{nonConse} becomes an operatorial identity (in the Heisenberg picture), while \eqref{hisguz} becomes a Kubo formula for the quantum statistical average $-u_\nu \braket{\hat{p}^\nu}$. Equation \eqref{worldlinecons} remains valid, if we interpret $G_0$ as the quantum statistical average $\braket{\hat{G}}_{\text{eq}}$, see appendix \ref{Rigiduz}. No further assumption about the equilibrium density matrix needs to be invoked in the proof. For example, we do not need to assume it to be of Gibbs-like form \citep{Gogoglin2016}, because this might point towards a von Neumann interpretation of the entropy, leading us back to circularity issues.

As a final comment, we remark that the Unruh effect \citep{Unruh1976} disappears in the limit in which the accelerations are adiabatic. In fact, with a simple order of magnitude estimate (see appendix \ref{UnruhLui}), one can verify that 
\begin{equation}
    (\text{Unruh corrections}) \sim \dfrac{e^{-1/\epsilon}}{\epsilon} \, .
\end{equation}
This shows that the Unruh effect is non-perturbative in $\epsilon$: it decays to zero faster than any finite power of $\epsilon$.

\section{Variation of the mass induced by a kick: quantum mechanics approach}

The proof of \eqref{iltaskuzzo} given above, using a field theory approach, makes the role of condition (ii) manifest. However, it somehow hides the physical meaning of our result. Why does a small kick conserve (to the first order) the mass of a system of particles in equilibrium, while accelerating it? Why must it be that
\begin{equation}\label{dacapire}
    \delta (p^\nu p_\nu) \sim \epsilon^2  \quad \quad \text{whereas}  \quad \quad   \delta p^\nu \sim \epsilon \, ?
\end{equation}
In this section, we will show, with a simple quantum mechanical argument, that \eqref{dacapire} is a consequence of the mathematical structure of the Poincar\'e group. The argument is rigorously formulated within relativistic quantum mechanics \citep{Keister1991}, while the connection with quantum field theory is somehow heuristic. This makes the argument that follows probably less conclusive than the one outlined in the previous section, but it gives a deeper insight into the dynamical origin of \eqref{dacapire}.

\subsection{The mass spectrum of a finite body}\label{spectrumz}

For the total four-momentum $p^\nu$ to be finite, the body must be of finite size. But a completely isolated finite body in thermodynamic equilibrium must be self-bounded \citep{GavassinoTermometri}, otherwise it would eventually break up into smaller pieces in relative motion. It is well-known from ordinary quantum mechanics that bound states of many particles have a discrete mass spectrum (as we see, for example, in nuclear and atomic physics). The intuition behind this fact is that the degrees of freedom of a many-body system decouple into center of mass degrees of freedom plus internal degrees of freedom. Since, in a bound state, the particles cannot escape the conglomerate\footnote{For large objects, at $T\neq 0$, perfect confinement is almost impossible and some form of radiation is always emitted. For this reason, relativistic thermodynamics is an idealization, which becomes valid in the limit in which the surface of bodies plays the role of a perfect mirror, keeping all the energy inside.}, the internal degrees of freedom (which describe essentially the relative positions between the particles) are bounded and, hence, have discrete energy eigenvalues. Recalling that the rest mass is the energy measured in the rest frame (i.e. it is the Hamiltonian of the internal degrees of freedom, see \citealt{Keister1991}), the discreteness of the mass eigenvalues follows.

Let us see the mathematical implications of the argument above. Given that the space-time translation operators $\hat{p}^\nu$, computed from the Lagrangian density $\mathcal{L}_{\text{Body}}$, commute with each other, we can take, as basis of the Hilbert space of the body, some states
\begin{equation}\label{statespa}
    \ket{p^\nu,a} \, ,
\end{equation}
satisfying the eigenvalue equations
\begin{equation}\label{eigenvalueequationpnu}
 \hat{p}^\nu \ket{p^\nu,a} = p^\nu \ket{p^\nu,a} \, .
\end{equation}
The additional quantum number $a$ is arbitrary (it is used to break possible degeneracies) and can be taken discrete. The eigenvalues $p^\nu$ must be continuous (they organise themselves into three-dimensional hyperboloids), due to the mathematical structure of the Poincar\'e group \citep{Weinberg_QFT_1995}. The square mass operator 
\begin{equation}
   \hat{M}^2 := -\hat{p}^\nu \hat{p}_\nu 
\end{equation}
commutes with all the generators of the Poincar\'e group (computed from $\mathcal{L}_{\text{Body}}$) and is diagonal on the basis \eqref{statespa}, with eigenvalue equation
\begin{equation}\label{autodimassa}
  \hat{M}^2  \ket{p^\nu,a} = m^2 \ket{p^\nu,a}  \spc m^2 =-p^\nu p_\nu \, .
\end{equation}
The scalar $m>0$ can be interpreted as the mass of the state $\ket{p^\nu,a}$. Combining the fact that $p^\nu$ is ``3D-continuous'', with the fact that $m$ and $a$ are discrete, we can conclude that  
\begin{equation}\label{normalizzo}
\braket{ \tilde{p}^\nu , \tilde{a}|p^\nu,a} = 2 p^0 (2\pi)^3 \delta^{(3)} (\tilde{p}^j - p^j) \, \delta_{\tilde{m} \, m}  \, \delta_{\tilde{a} \, a} \, .
\end{equation}
The standard normalization factor $2 p^0 (2\pi)^3$ guarantees that \eqref{normalizzo} is Lorentz-invariant \citep{Peskin_book}. 

Equation \eqref{normalizzo} is crucial for us, because it shows that we can build a \textit{normalisable} (i.e. physical) state $\ket{\Psi}$ which is eigenvector of the mass operator, namely
\begin{equation}\label{eigenf}
    \hat{M}\ket{\Psi} = m \ket{\Psi}.
\end{equation}
However, the same is not true for the individual components $\hat{p}^\nu$: the physical state $\ket{\Psi}$ must be a wavepacket, namely a continuous superposition of eigenstates of $\hat{p}^\nu$. As we are going to show, this is the central difference between $\hat{p}^\nu \hat{p}_\nu$ and $\hat{p}^\nu$, which is responsible for the different scalings of the corresponding perturbations.

\subsection{Kicking mass eigenstates}\label{kickingEigents}

Due to the presence of the term $\epsilon \, \phi \, G$ in the action \eqref{Action}, the operators $\hat{p}^\nu$ (which are computed from $\mathcal{L}_{\text{Body}}$) are not conserved during the kick. As a first step, let us compute the variation of the mass of the body, induced by a kick, when the initial state $\ket{\Psi}$ is an eigenvector of $\hat{M}$, satisfying the eigenvalue equation \eqref{eigenf}.

As $\phi(x^\nu)$ is an assigned function of the coordinates, the evolution of the body is unitary (the final state is still a pure state); this is the definition of thermal isolation \citep{landau5} or, equivalently, of no heat transfer \citep{Jaynes1965}. Working in the  Schr\"{o}dinger picture, we may call $\ket{\Psi_\epsilon(\tau)}$ the state of the body at the time $\tau$ (just after the perturbation has been switched off) as a function of the coupling constant $\epsilon$, parameterizing the intesity of the kick. Clearly, for $\epsilon=0$, the mass is conserved (no kick has occurred), so that we may write
\begin{equation}
    \delta M(\epsilon) = \dfrac{\braket{\Psi_\epsilon(\tau)|\hat{M}|\Psi_\epsilon(\tau)}}{\braket{\Psi_\epsilon(\tau)|\Psi_\epsilon(\tau)}} -\dfrac{\braket{\Psi_0(\tau)|\hat{M}|\Psi_0(\tau)}}{\braket{\Psi_0(\tau)|\Psi_0(\tau)}} \, .
\end{equation}
Expanding this function to the first order in $\epsilon$ we obtain
\begin{equation}
    \delta M(\epsilon) = \epsilon \dfrac{d}{d\epsilon} \bigg( \dfrac{\braket{\Psi_\epsilon(\tau)|\hat{M}|\Psi_\epsilon(\tau)}}{\braket{\Psi_\epsilon(\tau)|\Psi_\epsilon(\tau)}} \bigg) \bigg|_{\epsilon=0} + \mathcal{O}(\epsilon^2) \, .
\end{equation}
If we compute the derivative in $\epsilon$ explicitly, we get
\begin{equation}
   \delta M(\epsilon) = \epsilon \dfrac{\braket{\Psi'|\Delta} +\braket{\Delta|\Psi'}}{\braket{\Psi_0(\tau)|\Psi_0(\tau)}} + \mathcal{O}(\epsilon^2) \, ,
\end{equation}
with 
\begin{equation}\label{gringuzZo}
    \begin{split}
        & \ket{\Psi'} = \dfrac{d \ket{\Psi_\epsilon (\tau)}}{d  \epsilon} \bigg|_{\epsilon=0}  \\
        & \ket{\Delta} = \hat{M} \ket{\Psi_0(\tau)} - \dfrac{\braket{\Psi_0(\tau)|\hat{M}|\Psi_0(\tau)}}{\braket{\Psi_0(\tau)|\Psi_0(\tau)}} \ket{\Psi_0(\tau)} \, .  \\
    \end{split}
\end{equation}
The final step consists of realising that, if the initial state $\ket{\Psi}$ obeys equation \eqref{eigenf}, then 
\begin{equation}\label{eigenf2}
    \hat{M}\ket{\Psi_0 (\tau)} = m \ket{\Psi_0 (\tau)} \, ,
\end{equation}
because, when $\epsilon=0$, the Hamiltonian is $\hat{p}^0$, which commutes with $\hat{M}$. Inserting \eqref{eigenf2} into the second equation of \eqref{gringuzZo} we find $\ket{\Delta}=0$, which immediately implies
\begin{equation}\label{felixfelicis}
    \delta M (\epsilon) \sim \epsilon^2 \, .
\end{equation}
It is interesting to note that this result does not depend on the details of the full Hamiltonian of the system, because the explicit formula for $\ket{\Psi'}$ is completely irrelevant. However, the assumption that $\ket{\Psi}$ is a mass eigenstate is crucial. If we repeat the calculations above, taking as initial state a superposition
\begin{equation}
   \dfrac{ \ket{m_1} + \ket{m_2}}{\sqrt{2}} \, ,
\end{equation}
$\ket{m_1}$ and $\ket{m_2}$ being two normalised mass eingentates, relative to two different eigenvalues $m_1$ and $m_2$, we now obtain (truncating to the first order in $\epsilon$)
\begin{equation}\label{ghgh}
   \ket{\Delta} = \dfrac{m_1 -m_2}{2\sqrt{2}} \bigg(\ket{m_{1}(\tau)}-\ket{m_{2}(\tau)} \bigg)
\end{equation}
which does not vanish. By analogy, it becomes immediately clear why, in a kick, one is always able to induce an acceleration: any physical state must be a superposition of eigenstates of $\hat{p}^\nu$, hence the variation of $\braket{\Psi|\hat{p}^\nu|\Psi}$ is of order $\epsilon$ for the same reason why the variation of $(\bra{m_1}+ \bra{m_2}) \hat{M} (\ket{m_1}+\ket{m_2})$ is of order $\epsilon$.

\subsection{Kicking thermal states}

As we explained qualitatively in subsection \ref{classonz} (and proved rigorously in appendix \ref{Rigiduz}), a system that is in thermodynamic equilibrium has constant shape. Its internal structure is conserved over time and the only change that the system can experience is a rigid macroscopic motion. Given that $\hat{M}$ is the Hamiltonian of the internal degrees of freedom, it immediately follows that the density matrix of a macroscopic body in equilibrium satisfies the equation
\begin{equation}\label{commutazia}
    \Big[ \, \hat{\rho}_{\text{eq}} \, , \, \hat{M} \, \Big] =0 \, .
\end{equation}
It is not hard to show that this condition is essentially equivalent to equation \eqref{mountains} of appendix \ref{Rigiduz}.\footnote{There is a small difference between \eqref{commutazia} and \eqref{mountains}, which is due to the fact that the present description is entirely quantum-based, while in the appendix we use a hybrid approach. For macroscopic bodies this difference becomes negligible and the final result is the same.}

Equation \eqref{commutazia} implies that there is an orthonormal set of mass eigenstates $\ket{\Psi^{(n)}}$, with
\begin{equation}
    \hat{M}\ket{\Psi^{(n)}} = m_n \ket{\Psi^{(n)}}  \spc \braket{\Psi^{(\tilde{n})} | \Psi^{(n)}} = \delta_{\tilde{n} \, n} \, ,
\end{equation}
such that 
\begin{equation}
\hat{\rho}_{\text{eq}} = \sum_n \mathcal{P}_n \ket{\Psi^{(n)}} \bra{\Psi^{(n)}} \, 
\end{equation}
with
\begin{equation}
    \mathcal{P}_n >0 \spc  \sum_n \mathcal{P}_n =1   \, .
\end{equation}
Taking this as initial state and recalling that the evolution is unitary, it follows that the average value of $\hat{M}$ at a time $\tau$ (at the end of the kick) is
\begin{equation}
    M_\tau = \sum_n \mathcal{P}_n \dfrac{\bra{\Psi^{(n)}_\epsilon(\tau)} \hat{M}\ket{\Psi^{(n)}_\epsilon(\tau)}}{\braket{\Psi^{(n)}_\epsilon(\tau) |\Psi^{(n)}_\epsilon(\tau)}}  \, .
\end{equation}
Given that equation \eqref{felixfelicis} applies to each contribution in the sum over $n$ (because each state $\ket{\Psi^{(n)}_\epsilon(\tau)}$ is the time-evolved of a mass eigenstate), it applies also  to  a  body  with density matrix $\hat{\rho}_{\text{eq}}$, completing our proof.

There is a final remark that we need to make. All our analysis was performed within the assumption that the system does not radiate particles as a result of the kick (particles can be created and destroyed \textit{inside the body}, but no particle can abandon the body). This is an important assumption, because, if it happens that the system emits particles along the way, the calculations above remain valid, but the quantity $M$ can no longer be interpreted as the mass of the body alone, but as the rest-frame energy of the \textit{total system} (``$\, \text{body}+\text{emitted particles} \,$'') invalidating the assumptions that lead to \eqref{dse}. Luckily, one can easily prove (see appendix \ref{emmizuz}) that also the probability of stimulated emissions induced by a kick is of the order $\epsilon^2$ (and, therefore, vanishes for adiabatic accelerations).

\section{Conclusions}

We have proved that the equation of state of isolated moving bodies (including only the four-momentum among the relevant variables) is always $S=S(M)$. Rather than showing this by arbitrarily postulating the Lorentz covariance of the laws of thermodynamics, we have focused on the dynamical consequence of assuming $S=S(M)$. In fact, declaring that two macroscopic states $\psi$ and $\psi'$ have the same entropy is equivalent to stating that there must be an adiabatic transformation that leads from $\psi$ to $\psi'$ and vice-versa. Using tools from both classical and quantum field theory we have shown that, indeed, infinitely slow accelerations, generated by a time-dependence of the Hamiltonian, must conserve the rest mass of bodies initially in thermodynamic equilibrium, making $S=S(M)$ the only equation of state possible. 

This sets solid foundations for relativistic thermodynamics and, again, shows that the axiomatization proposed by \cite{vanKampen1968} and \cite{Israel2009_book} is the only one possible. Furthermore, this paper complements our previous study on the nature of the temperature \citep{GavassinoTermometri}, in that it clarifies further the meaning of the work four-vector $\delta \mathcal{W}^\nu$. In the same way in which one may intuitively decompose the heat four-vector $\delta \mathcal{Q}^\nu$ into time and space components as 
\begin{equation}
    \delta \mathcal{Q}^\nu = \binom{\text{ Heat }}{\text{ Friction }} \, ,
\end{equation}
one may consider the analogous (non-rigorous but useful) pictorial decomposition of the work four-vector as 
\begin{equation}
    \delta \mathcal{W}^\nu = \binom{\text{ Work }}{\text{ Kick }} \, .
\end{equation}
In the same way in which a wall can exert work on a gas (changing its energy), a potential can induce a kick on a freely moving body (changing its momentum). Both these processes, if executed slowly enough, become reversible. The first becomes the standard pressure-volume (``PdV'') adiabatic work, while the second becomes a pure acceleration.

\section*{Acknowledgements}

 The author thanks M. Antonelli and B. Haskell for reading the manuscript and providing critical comments. I am particularly grateful to Prof. G. Torrieri, for his insights into the particle emission problem. I acknowledge support from the Polish National Science Centre grant OPUS 2019/33/B/ST9/00942. Partial support comes from PHAROS, COST Action CA16214.

\appendix

\section{Planck's original argument revisited}\label{AAA}

Here we present a new version of Planck's original argument \citep{Planck_1908}, which is slightly less abstract, but logically equivalent. 

Let $S_X$ be the entropy of a body $X$, as measured in its own rest frame, and assume that, if an observer $A$ (say, Alice) sees $X$ moving with a given speed $v_{XA}$, she will attribute to $X$ an entropy 
\begin{equation}
S_A = \gamma^n_{XA} S_X  \spc  \gamma_{XA}= (1-v_{XA}^2)^{-1/2},
\end{equation} 
with $n$ a constant exponent to be determined. This is a reasonably general assumption about the transformation law of the entropy, as it includes the possibility for the entropy to be a scalar ($n=0$), the zeroth component of a four-vector ($n=1$) or an arbitrary power of it.

Let us consider a second observer $B$ (say, Bob), in motion with respect to Alice. Bob should assign an entropy to $X$ using a rule that is analogous to that of Alice (there is nothing spacial about Alice's frame), namely
\begin{equation}
S_B = \gamma^n_{XB} S_X  \spc  \gamma_{XB}= (1-v_{XB}^2)^{-1/2},
\end{equation} 
where $v_{XB}$ is the speed of $X$ with respect to Bob. Now, assume that $X$ is initially at rest with respect to Alice (namely, $\gamma_{XA}=1$) and consider an infinitesimal reversible transformation in which $X$ is slowly set into motion. If the transformation is reversible, it should conserve the entropy in Alice's reference frame, hence
\begin{equation}\label{fromAlice}
0=\delta S_A = \delta (\gamma^n_{XA} S_X ) = \gamma^n_{XA} \delta S_X = \delta S_X \, ,
\end{equation}
where we have used the fact that
\begin{equation}
\delta \gamma_{XA} = \gamma_{XA}^3 v_{XA}\delta v_{XA} =0,
\end{equation}
because initially $v_{XA}=0$. On the other hand, it should be reversible also in the reference frame of Bob, because reversibility is a statement about the possibility of both the process itself and its inverse to occur, which cannot depend on the observer. Hence, using \eqref{fromAlice}, we obtain
\begin{equation}
0=\delta S_B = \delta (\gamma^n_{XB} S_X ) = n S_X \gamma^{n+2}_{XB} v_{XB} \delta v_{XB}.
\end{equation}
Considering that in general $ v_{XB} \, \delta v_{XB}\neq 0$, the exponent $n$ must vanish and, consequently, the entropy must be a scalar.

\section{Isolated bodies in equilibrium move rigidly}\label{Rigiduz}

Here we prove that, when a body in thermodynamic equilibrium is perfectly isolated (also dynamically, hence $\epsilon =0$), it drifts rigidly at constant velocity without experiencing any change of shape. The key assumption to be used in the proof is condition (ii), see subsection \ref{assumptionsSSSSS}. We recall that all the calculation are performed, for clarity, in the preferred reference frame $A$ introduced in condition (i).

\subsection{Centroids}

It is always possible to build, starting from the Noether stress-energy tensor $T^{\nu \rho}$, which is not necessarily symmetric, the Belinfante-Rosenfeld stress-energy tensor $\Theta^{\nu \rho}$, which is symmetric \citep{Weinberg_QFT_1995}. This tensor field can be used to write the angular momentum tensor $J^{\nu \rho}$, which is the generator of the Lorentz group and is conserved (for isolated bodies), in the form 
\begin{equation}
    J^{\nu \rho} = \int \big( x^\nu \Theta^{\rho 0} - x^\rho \Theta^{\nu 0} \big) \, d^3 x \, .
\end{equation}
This formula can be used to show that the conservation of $J^{0j}$ implies \citep{MTW_book}
\begin{equation}
    \dfrac{dx_{CD}^j}{dt} = \dfrac{p^j}{p^0} = \dfrac{u^j}{u^0} =: v^j \, ,
\end{equation}
where $x_{CD}^j$ is the position of the centroid of the system, defined as
\begin{equation}
    x_{CD}^j = \dfrac{1}{p^0} \int x^j \Theta^{00} \, d^3 x \, .
\end{equation}
This well-known fact is the relativistic generalization of the Newtonian law according to which the center of mass evolves following a uniform rectilinear motion. No matter how complicated the internal dynamics of the body is, as long as the body is isolated, condition (ii) guarantees that
\begin{equation}\label{IlCentroide}
    x_{CD}^j(t) = x_{CD}^j(0) +v^j t \, . 
\end{equation}

\subsection{The density matrix at equilibrium}

Consider a body in equilibrium with four-momentum $p^\nu$ and centroid $x_{CD}^j$, at a given time. The state of the system must be a function of these parameters. 
Since in quantum mechanics any physical state can be modelled through a density matrix, there must be a formula
\begin{equation}\label{densuz}
    \hat{\rho}_{\text{eq}}[p^\nu , x_{CD}^j] \, ,
\end{equation}
which gives all the physical properties of a system in thermodynamic equilibrium as a function of its four-momentum $p^\nu$ and of its centroid $x_{CD}^j$. No other free parameter needs to be included in \eqref{densuz} because we are assuming that there are no (relevant) additional constants of motion. We do not specify any precise formula for \eqref{densuz}, because this would imply giving a statistical interpretation to the entropy, which is something we want to avoid here. 

If $\hat{p}^\nu$ are the four-momentum operators, then the unitary operator
\begin{equation}
    \hat{U}(\Delta x^\nu) = \exp{\big(-i\hat{p}_\nu \, \Delta x^\nu \big)}
\end{equation}
is a space-time translation, which acts on the field operators $\hat{\varphi}_i$ as follows \citep{Weinberg_QFT_1995}:
\begin{equation}
 \hat{U}(\Delta x^\nu) \, \hat{\varphi}_i (x^\nu) \, \hat{U}^\dagger(\Delta x^\nu) = \hat{\varphi}_i (x^\nu +\Delta x^\nu) \, .
\end{equation}
Clearly, if we consider a system in thermodynamic equilibrium, and we operate on it a pure translation in space, the final state must still be an equilibrium state, namely
\begin{equation}\label{trasliamoinsieme}
  \hat{U}(\Delta x^j) \, \hat{\rho}_{\text{eq}}[p^\nu , x_{CD}^j] \, \hat{U}^\dagger(\Delta x^j) = \hat{\rho}_{\text{eq}}[p^\nu , x_{CD}^j+\Delta x^j] \, .
\end{equation}

\subsection{Rigid motion}

By definition, if an isolated body is in equilibrium at a given time, it is also in equilibrium at later times. Hence, recalling that the four-momentum is conserved and that equation \eqref{IlCentroide} must hold, we have
\begin{equation}
    e^{-i \hat{p}^0 \Delta t} \, \hat{\rho}_{\text{eq}}[p^\nu , x_{CD}^j] \,  e^{i \hat{p}^0 \Delta t} = \hat{\rho}_{\text{eq}}[p^\nu , x_{CD}^j+v^j \Delta t] \, . 
\end{equation}
Using equation \eqref{trasliamoinsieme}, we find
\begin{equation}\label{mountains}
    \hat{\rho}_{\text{eq}}[p^\nu , x_{CD}^j] = \hat{U} \bigg( \dfrac{u^\nu \Delta t}{u^0} \bigg) \, \hat{\rho}_{\text{eq}}[p^\nu , x_{CD}^j] \,  \hat{U}^\dagger\bigg( \dfrac{u^\nu \Delta t}{u^0} \bigg) . 
\end{equation}
Multiplying this equation by an arbitrary field of observables $\hat{G}(x^\nu)$ and taking the trace of the result, we find
\begin{equation}
    \text{tr} \bigg[ \hat{\rho}_{\text{eq}} \hat{G}(x^\nu) \bigg] = \text{tr} \bigg[ \hat{\rho}_{\text{eq}} \hat{G}\bigg(x^\nu - \dfrac{u^\nu \Delta t}{u^0} \bigg) \bigg] \, .
\end{equation}
Taking the derivative of this formula with respect to $\Delta t$ we finally obtain
\begin{equation}\label{Laprovaprovata}
    u^\nu \partial_\nu \braket{\hat{G}}_{\text{eq}}=0 \, ,
\end{equation}
which is what we wanted to prove.

Note that this result is in perfect agreement with the relativistic formulation of the zeroth law of thermodynamics. In fact, following \cite{GavassinoTermometri}, in thermal equilibrium there must be one reference frame in which perfect stationarity is achieved. From \eqref{Laprovaprovata}, we see that this reference frame is identified by the four-velocity $u^\nu=p^\nu/M$, in agreement with \cite{GavassinoTermometri}.

\section{Effect of absorption/emission processes on adiabatic accelerations}

In \ref{UnruhLui} we show that the Unruh effect does not play any role in adiabatic accelerations. In \ref{emmizuz} we show that the probability of emission of particles out of the body, stimulated by the kick, is of order $\epsilon^2$.

\subsection{Unruh effect for adiabatic accelerations}\label{UnruhLui}

From the point of view of a particle detector that accelerates with constant acceleration $a$, the average number of particles (scalar bosons, for simplicity) with energy $E$ is given by \citep{UnruhWald1984,LinHu2006}
\begin{equation}
    f(E) = \dfrac{1}{e^{2\pi E/a}-1} \, .
\end{equation}
The detector can make level transitions by absorbing and emitting particles, with a stimulated absorption/emission rate which is proportional to $f(\Delta m)$, where $\Delta m >0$ is the mass separation between two levels of the detector. Therefore, if a detector experiences a uniform acceleration $a$, for an interval of proper time $t$, the transition probabilities (associated with Unruh-particle absorption/emission processes) scale as \citep{LinHu2007}
\begin{equation}
    \mathcal{P}_{Unruh} \sim \dfrac{t}{e^{2\pi \Delta m/a}-1} \, .
\end{equation}
For a reversible acceleration, as described in subsection \eqref{infinite infitesimal}, $t \sim N$, while $a \sim 1/N$, hence
\begin{equation}
    \mathcal{P}_{Unruh} \sim \dfrac{N}{e^{N}-1} \sim N e^{-N} \longrightarrow 0 .
\end{equation}
Therefore, in the limit of adiabatic accelerations ($N \rightarrow +\infty$), the corrections due to the Unruh effect are exponentially suppressed.

\subsection{Can an adiabatic acceleration stimulate emissions?}\label{emmizuz}

The Hilbert space of the body, generated by the discrete-mass basis $ \ket{p^\nu,a}$, introduced in equation \eqref{statespa}, is only a subspace of the full Hilbert space upon which the quantum fields $\hat{\varphi}_i$ act (as operators). There are many other states, including, in particular, states in which the body coexists with other particles. These states constitute the continuous part of the mass spectrum \citep{Peskin_book}. The projector
\begin{equation}
    \hat{P} := \hat{\mathbb{I}} - \sum_{m,a} \int \dfrac{d^3p}{(2\pi)^3} \dfrac{\ket{p^\nu ,a} \bra{p^\nu,a}}{2p^0} 
\end{equation}
projects onto this second part of the Hilbert space ($\hat{\mathbb{I}}$ is the identity operator acting on total Hilbert space of the field theory). Given a normalised state $\ket{\Psi}$, the average $\bra{\Psi} \hat{P} \ket{\Psi}$ is the probability that we observe ``something that is not just the body alone''. Therefore, we can interpret the quantum average
\begin{equation}
    \mathcal{P}_{\text{Em}}(\tau) = \text{tr} \bigg[ \hat{\rho}(\tau)  \hat{P} \bigg] 
\end{equation}
as the probability that the body has emitted something during a kick. Now, from equation \eqref{eigenvalueequationpnu} it follows that (as long as $\epsilon=0$)
\begin{equation}\label{aszumo}
     \Big[ \, \hat{P} \, , \, \hat{p}^0 \, \Big] =0 \, .
\end{equation}
Furthermore, since in the initial state there is only the body, we know that
\begin{equation}\label{aszumo2}
     \hat{\rho}(0) \hat{P} = 0.
\end{equation}
Combining \eqref{aszumo} with \eqref{aszumo2} we immediately find that, if $\epsilon=0$, $\mathcal{P}_{\text{Em}}(\tau)$ vanishes. On the other hand, $\mathcal{P}_{\text{Em}}(\tau) \geq 0$ (recall that $\hat{P}=\hat{P}^\dagger=\hat{P}^2$), therefore we cannot impose $\mathcal{P}_{\text{Em}}(\tau) \sim \epsilon$, because $\epsilon$ has arbitrary sign. Hence, the leading order must be
\begin{equation}
    \mathcal{P}_{\text{Em}}(\tau) \sim \epsilon^2 \, , 
\end{equation}
or higher (but even), which is what we wanted to prove.

\bibliographystyle{mnras}
\bibliography{Biblio}

\begin{thebibliography}{}
\makeatletter
\relax
\def\mn@urlcharsother{\let\do\@makeother \do\$\do\&\do\#\do\^\do\_\do\%\do\~}
\def\mn@doi{\begingroup\mn@urlcharsother \@ifnextchar [ {\mn@doi@}
  {\mn@doi@[]}}
\def\mn@doi@[#1]#2{\def\@tempa{#1}\ifx\@tempa\@empty \href
  {http://dx.doi.org/#2} {doi:#2}\else \href {http://dx.doi.org/#2} {#1}\fi
  \endgroup}
\def\mn@eprint#1#2{\mn@eprint@#1:#2::\@nil}
\def\mn@eprint@arXiv#1{\href {http://arxiv.org/abs/#1} {{\tt arXiv:#1}}}
\def\mn@eprint@dblp#1{\href {http://dblp.uni-trier.de/rec/bibtex/#1.xml}
  {dblp:#1}}
\def\mn@eprint@#1:#2:#3:#4\@nil{\def\@tempa {#1}\def\@tempb {#2}\def\@tempc
  {#3}\ifx \@tempc \@empty \let \@tempc \@tempb \let \@tempb \@tempa \fi \ifx
  \@tempb \@empty \def\@tempb {arXiv}\fi \@ifundefined
  {mn@eprint@\@tempb}{\@tempb:\@tempc}{\expandafter \expandafter \csname
  mn@eprint@\@tempb\endcsname \expandafter{\@tempc}}}

\bibitem[\protect\citeauthoryear{{Adami}}{{Adami}}{2011}]{Adami2011}
{Adami} C.,  2011, arXiv e-prints, \href
  {https://ui.adsabs.harvard.edu/abs/2011arXiv1112.1941A} {p. arXiv:1112.1941}

\bibitem[\protect\citeauthoryear{{Becattini}}{{Becattini}}{2016}]{Becattini2016}
{Becattini} F.,  2016, \mn@doi [Acta Physica Polonica B]
  {10.5506/APhysPolB.47.1819}, \href
  {https://ui.adsabs.harvard.edu/abs/2016AcPPB..47.1819B} {47, 1819}

\bibitem[\protect\citeauthoryear{{Cercignani} \& {Kremer}}{{Cercignani} \&
  {Kremer}}{2002}]{cercignani_book}
{Cercignani} C.,  {Kremer} G.~M.,  2002, {The relativistic Boltzmann equation:
  theory and applications}

\bibitem[\protect\citeauthoryear{De~Groot}{De~Groot}{1980}]{degroot_book}
De~Groot S.,  1980, {Relativistic Kinetic Theory. Principles and Applications}

\bibitem[\protect\citeauthoryear{Farias, Pinto  \& Moya}{Farias
  et~al.}{2017}]{Farias2017}
Farias C.,  Pinto V.,   Moya P.,  2017, \mn@doi [Scientific Reports]
  {10.1038/s41598-017-17526-4}, 7, 17657

\bibitem[\protect\citeauthoryear{{Gavassino}}{{Gavassino}}{2020}]{GavassinoTermometri}
{Gavassino} L.,  2020, \mn@doi [Foundations of Physics]
  {10.1007/s10701-020-00393-x}, \href
  {https://ui.adsabs.harvard.edu/abs/2020FoPh..tmp..117G} {}

\bibitem[\protect\citeauthoryear{{Gavassino}, {Antonelli}  \&
  {Haskell}}{{Gavassino} et~al.}{2021}]{GavassinoIorda2021}
{Gavassino} L.,  {Antonelli} M.,   {Haskell} B.,  2021, \mn@doi [Universe]
  {10.3390/universe7020028}, \href
  {https://ui.adsabs.harvard.edu/abs/2021Univ....7...28G} {7, 28}

\bibitem[\protect\citeauthoryear{{Gogolin} \& {Eisert}}{{Gogolin} \&
  {Eisert}}{2016}]{Gogoglin2016}
{Gogolin} C.,  {Eisert} J.,  2016, \mn@doi [Reports on Progress in Physics]
  {10.1088/0034-4885/79/5/056001}, \href
  {https://ui.adsabs.harvard.edu/abs/2016RPPh...79e6001G} {79, 056001}

\bibitem[\protect\citeauthoryear{Hakim}{Hakim}{2011}]{hakim_book}
Hakim R.,  2011, {Introduction to Relativistic Statistical Mechanics: Classical
  and Quantum}.
World Scientific

\bibitem[\protect\citeauthoryear{Huang}{Huang}{1987}]{huang_book}
Huang K.,  1987, Statistical Mechanics, 2 edn.
John Wiley \& Sons

\bibitem[\protect\citeauthoryear{Israel}{Israel}{1981}]{Israel_1981_review}
Israel W.,  1981, \mn@doi [Physica A: Statistical Mechanics and its
  Applications] {https://doi.org/10.1016/0378-4371(81)90220-X}, 106, 204

\bibitem[\protect\citeauthoryear{Israel}{Israel}{2009}]{Israel2009_book}
Israel W.,  2009, Relativistic Thermodynamics.
Birkh{\"a}user Basel, Basel, pp 101--113, \mn@doi{10.1007/978-3-7643-8878-2_8},
  \url {https://doi.org/10.1007/978-3-7643-8878-2_8}

\bibitem[\protect\citeauthoryear{Israel \& Stewart}{Israel \&
  Stewart}{1979}]{Israel_Stewart_1979}
Israel W.,  Stewart J.,  1979, \mn@doi [Annals of Physics]
  {https://doi.org/10.1016/0003-4916(79)90130-1}, 118, 341

\bibitem[\protect\citeauthoryear{{Jaynes}}{{Jaynes}}{1965}]{Jaynes1965}
{Jaynes} E.~T.,  1965, \mn@doi [American Journal of Physics]
  {10.1119/1.1971557}, \href
  {https://ui.adsabs.harvard.edu/abs/1965AmJPh..33..391J} {33, 391}

\bibitem[\protect\citeauthoryear{Keister \& Polyzou}{Keister \&
  Polyzou}{1991}]{Keister1991}
Keister B.~D.,  Polyzou W.~N.,  1991, Adv. Nucl. Phys., 20, 225

\bibitem[\protect\citeauthoryear{Landau \& Lifshitz}{Landau \&
  Lifshitz}{2013}]{landau5}
Landau L.,  Lifshitz E.,  2013, Statistical Physics.
No.~v. 5, Elsevier Science, \url
  {https://books.google.pl/books?id=VzgJN-XPTRsC}

\bibitem[\protect\citeauthoryear{Lin \& Hu}{Lin \& Hu}{2006}]{LinHu2006}
Lin S.-Y.,  Hu B.~L.,  2006, \mn@doi [Phys. Rev. D]
  {10.1103/PhysRevD.73.124018}, 73, 124018

\bibitem[\protect\citeauthoryear{{Lin} \& {Hu}}{{Lin} \&
  {Hu}}{2007}]{LinHu2007}
{Lin} S.-Y.,  {Hu} B.~L.,  2007, \mn@doi [\prd] {10.1103/PhysRevD.76.064008},
  \href {https://ui.adsabs.harvard.edu/abs/2007PhRvD..76f4008L} {76, 064008}

\bibitem[\protect\citeauthoryear{{Mare{\v{s}}}, Hub\'{i}k  \&
  Spicka}{{Mare{\v{s}}} et~al.}{2017}]{Mares2017}
{Mare{\v{s}}} J.,  Hub\'{i}k P.,   Spicka V.,  2017, \mn@doi [Fortschritte der
  Physik] {10.1002/prop.201700018}, 65, 1700018

\bibitem[\protect\citeauthoryear{{Misner}, {Thorne}  \& {Wheeler}}{{Misner}
  et~al.}{1973}]{MTW_book}
{Misner} C.~W.,  {Thorne} K.~S.,   {Wheeler} J.~A.,  1973, {Gravitation}

\bibitem[\protect\citeauthoryear{{Nakamura}}{{Nakamura}}{2012}]{Nakamura2012}
{Nakamura} T.~K.,  2012, Progress of Theoretical Physics, \href
  {https://ui.adsabs.harvard.edu/abs/2012PThPh.128..463N} {128, 463}

\bibitem[\protect\citeauthoryear{Parvan}{Parvan}{2019}]{PARVAN2019}
Parvan A.,  2019, \mn@doi [Annals of Physics]
  {https://doi.org/10.1016/j.aop.2019.01.003}, 401, 130

\bibitem[\protect\citeauthoryear{Peskin \& Schroeder}{Peskin \&
  Schroeder}{1995}]{Peskin_book}
Peskin M.~E.,  Schroeder D.~V.,  1995, {An Introduction to quantum field
  theory}.
Addison-Wesley, Reading, USA, \url
  {http://www.slac.stanford.edu/~mpeskin/QFT.html}

\bibitem[\protect\citeauthoryear{Planck}{Planck}{1908}]{Planck_1908}
Planck M.,  1908, \mn@doi [Annalen der Physik] {10.1002/andp.19083310602}, 331,
  1

\bibitem[\protect\citeauthoryear{{Rezzolla} \& {Zanotti}}{{Rezzolla} \&
  {Zanotti}}{2013}]{rezzolla_book}
{Rezzolla} L.,  {Zanotti} O.,  2013, {Relativistic Hydrodynamics}

\bibitem[\protect\citeauthoryear{{Rigol}, {Dunjko}  \& {Olshanii}}{{Rigol}
  et~al.}{2008}]{Rigol2008}
{Rigol} M.,  {Dunjko} V.,   {Olshanii} M.,  2008, \mn@doi [\nat]
  {10.1038/nature06838}, \href
  {https://ui.adsabs.harvard.edu/abs/2008Natur.452..854R} {452, 854}

\bibitem[\protect\citeauthoryear{Unruh}{Unruh}{1976}]{Unruh1976}
Unruh W.~G.,  1976, \mn@doi [Phys. Rev. D] {10.1103/PhysRevD.14.870}, 14, 870

\bibitem[\protect\citeauthoryear{Unruh \& Wald}{Unruh \&
  Wald}{1984}]{UnruhWald1984}
Unruh W.~G.,  Wald R.~M.,  1984, \mn@doi [Phys. Rev. D]
  {10.1103/PhysRevD.29.1047}, 29, 1047

\bibitem[\protect\citeauthoryear{{Weinberg}}{{Weinberg}}{1995}]{Weinberg_QFT_1995}
{Weinberg} S.,  1995, {The Quantum Theory of Fields Volume I: Foundations}

\bibitem[\protect\citeauthoryear{van Kampen}{van Kampen}{1968}]{vanKampen1968}
van Kampen N.~G.,  1968, \mn@doi [Phys. Rev.] {10.1103/PhysRev.173.295}, 173,
  295

\makeatother
\end{thebibliography}

\end{document}